Ising Matter as Bulky: An Identity Described by Many Models

Martin H. Krieger, Krieger@usc.edu University of Southern California, Los Angele 90089-0626

Abstract

The two-dimensional Ising model of a ferromagnet allows for many ways of computing its partition function and other properties. Each way reveals surprising features of what we might call "Ising matter." Moreover, the various ways would appear to analogize with mathematicians' threefold analogy of analysis, algebra, and arithmetic, due to R. Dedekind and H. Weber, 1882, and more recently described by A. Weil.

Modeling Bulkiness An Identity Described by Many Models[1]

"Bulk matter" is what we experience everyday in a stone, in a flowing stream of water, a piece of wood, an ice cube, a magnet. The Ising lattice--imagine a two-dimensional grid, effectively infinite, square, with magnetic spins at each vertex, each interacting with its nearest neighbors--as a kind of bulk matter, may be seen and understood in manifold ways (different qualities, models, presentations or profiles or perspectives). Yet, the ways all are about that same lattice, that matter, that "identity," and the ways are complementary even though they might well be very different in style and substance.

When physical scientists provide an account of everyday Nature, they think of matter in a variety of ways. Fundamentally, there are the atoms and molecules interacting with each other and with external fields (such as a magnetic field), sometimes attractively, sometimes repulsively, sometimes by collision, sometimes at a (small) distance. But that matter might be a gas, a liquid, or a solid--a bulk that has its own properties and symmetries. That bulk matter may be crystalline, it may allow for flows and turbulence, it may allow for rigidity and stress and strain. It may be of fixed volume, or able to fill any volume. And likely it has a temperature, a volume, etc.

Along the way, physicists might think of that matter as composed of molecules whose total effect on each other, one-by-one, is, "adds up to," that bulk matter. Or, the molecules might thought of as being linked to each other, as in a long open or closed chain (as in a polymer or a polygon), perhaps intertwined, there being many such chains, again leading to that bulk matter. Or, the molecules might be grouped into small collections, those small collections then grouped into slightly larger collections, and so forth, until we have something like bulk matter, in which, when we put together two such super-collections, two stones, for example, not much happens other than they add up, so to speak. Or, we might find that at whatever scale we look at (once the scale is somewhat larger than an intermolecular distance), matter looks the same at each scale. And, surely there are other such modes of constituting matter.[2]

The various constitutions of matter are chosen because they allow for perspicuous understanding of what is going on in Nature, and for theoretical and mathematical calculation of those phenomena from first principles. Each constitution is a model, and then one goes to work. One problem is to account for bulk matter given its molecular constitution. Might we show that ordinary everyday matter is stable, and does not explode or collapse into itself, and that it has reasonable properties, such as if we compress it, it becomes less voluminous (what is called "thermodynamic stability"). Another is to account for flows and turbulence of bulk matter. And another is to account for symmetries and structures, as in a crystalline solid.

Another problem is to account for the collective behavior of a large number of molecules, so that they appear to become a coherent whole. To repeat, one such problem is how matter might become solid, or liquid, or a gas, and another is to show how a solid might have an orderly crystalline structure. A third problem, the one I shall focus on here, is that of magnetism, that a bunch of (say, iron) molecules might align their magnetic spins so that we have a permanent magnet. Or, more generally, accounting for the transition from disorder at very high temperatures to order and bulk magnetization, a "permanent magnet," at lower temperatures. Here, the intermolecular attraction that would lead to the alignment of the spins competes with the random motion of the molecules due to their thermal energy, and the question is whether that attraction is ever able to defeat the thermal energy of random vibration and rotation sufficiently so that the molecules' spins are lined up.

To Recall The Ising Model

The Ising model of a two-dimensional magnetic material has magnetic spins at each of the vertices of a lattice, the spins influenced by their neighbors so that they might align (their interaction hamiltonian is $-J\sigma_i\sigma_{i+1}$, $\sigma=\pm1$) yet also influenced by their thermal energy that would make such alignment very difficult since each of the spins might rotate any which way in the plane (here just up or down). One might treat such matter as a collection of individual molecules, as polymeric, as blocks of blocks of molecules, or as exhibiting scaling symmetry. Each perspective on the constitution of Ising matter is only notional until we have an effective way of computing its properties—its thermodynamic free energy, its transition from disorder to order as the temperature is lowered (below about 1000 degrees Kelvin, for iron) so its becoming a permanent magnet, its response to external magnetic fields.[3] Presumably all such ways should come up with the same answers, although each way will reveal particular aspects of such matter, aspects that might be difficult to discern from another perspective.

Mathematical Techniques for Solution

The mathematical techniques employed in solving the Ising model, that is, in computing its thermodynamic free energy or its permanent magnetization or its response to external magnetic fields are vertiginous. One computes what is called here the partition function, a weighted sum of the number of different configurations (each molecule either up or down) of the molecules in that bulk at a particular energy. The logarithm of that partition function is proportional to the chemists' Free Energy, $FE$ (Partition Function = $\exp -FE/k_BT$). The partition function's zeros, in the limit of bulk matter, indicate phase transitions. We might think of the partition function as a weighted averaging operator, averaging the number of states having a particular energy, over all energies. [4]

Each of these are well-defined mathematical techniques:

a. *Arithmetic or Combinatorial Counting-Up*: To compute the partition function, one might *count up* all the various configurations of all the possible up and down spins (a combinatorial problem), with a suitable weight of their probability (the Boltzmann factor, $\exp -Energy/k_BT$, the *Energy* of a state being defined by the sum of the energies of each of the pair interactions of neighboring molecules, $k_B$ being Boltzmann's constant, and $T$ is the Kelvin temperature). One might do so one pair at a time, maybe one row at a time, or perhaps one quadrant at a time. Or, one might account for Ising matter by defining suitable polygons of linked molecules and count them up, using notions from the theory of graphs (expressed as a Pfaffian of a matrix, the Pfaffian being much like a determinant that automatically counts up the interactions but now without double-counting). Or, perhaps, consider cluster expansions of groups of molecules, going from 2 to 3 to 4…interacting molecules, three molecules interacting in a two-by-two fashion, so that we have three such interactions.

b. *Counting-Up by Recognizing Symmetries*: One might discern some *symmetries* of the Ising lattice: a very high temperature lattice with a bit of order is like a very low temperature lattice with a bit of disorder; or, that as we add more disorder to a very orderly lattice, its properties scale according to something like its temperature; or, that a lattice that is triangular has properties not so different from a lattice that is hexagonal (as in the Wye-Delta transformation in electrical circuit theory). From such symmetries, one might discern an equation for the partition function. To solve this equation, one needs to make assumptions about the analyticity (smoothness, continuity) of its solutions, at least in some regions.

In fact there is a parameter, $k$, that indexes the temperature of the lattice. Close to its freezing or critical point, $k \approx 1$, so that $k-1$ is a measure of the distance to that critical point. And there is another parameter, $u$, that is relevant. $k$ and $u$ turn out to be the modulus and argument of the elliptic functions. The symmetry of high and low temperature lattices is a $k$ to $1/k$ transformation, the triangle-hexagonal is $k$-preserving but changes $u$. There are cousins of the sines and cosines (the circular or trigonometric functions), those elliptic functions (such as $sn$), and their behavior models what happens to the partition function—$sn(k)$ is proportional to $sn(1/k)$, namely, $sn(u,k) = 1/k \times sn(ku,1/k)$, when the high-low temperature transformation is made.

c. *Counting-Up by Means of Algebraic Devices*: One might develop an algebra that would account for each of the pair-interactions as a whole, in effect by-the-way doing the sum of *a*. Such an algebra turns out to be reminiscent of quantum mechanics, where again one wants to add up all the ways of making up a system, actually the paths to it, although this is a classical problem. Moreover, the partition function is here a trace of the crucial "transfer matrix," which automatically adds up all the interactions, in effect transferring down the line of spins picking up all the interaction energies along the way. Those matrices commute if they are characterized by the same $k$, and it would appear they are "group representations" of the symmetries of the Ising lattice.

*Avoiding the Arithmetic, Symmetries, and Algebra:*

d. *Counting-Up by Studying the Consequences of Doing So:* One might study *interfaces* of the system of molecules, interfaces like those polymers or polygons separating clusters of like-pointing up-molecules from clusters of like-pointing down-molecules. There will be islands of like-pointing molecules, and as the temperature lowers, those islands become larger, but all other size islands are also present. One is looking for an interface that in effect divides the lattice into an up-island and a down-island. One assumes that understanding this interface will be informative of the bulk properties of the Ising matter.[5]

Ingenious devices have been employed to do these calculations: Kac and Ward showing how the combinatorics of *a* (rather than the algebra of *c*) might be encoded in the determinant of a jerry-built matrix; Kramers-Wannier "duality" (the *k* to 1/*k* transformation); Baxter's legerdemain with symmetries and functional equations. But in time all of these tricks have become domesticated and incorporated into meaningful methods of solution, under the demands of physical insight and analogy, and mathematical generalization and rigor.

An analogy of analogies, between physics and mathematics, of analysis, algebra, and arithmetic

*c, b, a,* respectively, parallel the methods developed by Dedekind and Weber, 1882, for better understanding Riemann's geometrical and continuous picture of a surface that represents a function, such as the $\sqrt{x}$, which has multiple values.[6] In order to develop an *algebraic* (*c*) function theory (functions defined as roots of polynomial equations) that was seen to be needed, to make Riemann's *geometrical* insights (*b*) more manifestly rigorous, they used as a model Dedekind's "algebraic number theory" (*a*) (which accounts for the fact that some number systems, unlike our integers, do not have unique factorization, so if we add $\sqrt{5}$ to our number system, $6 = 2 \times 3 = (1+\sqrt{5}) \times (1-\sqrt{5})$).

In effect, the physicists in their solving the Ising model in so many ways would appear to reproduce, without being aware of it, this threefold analogy developed in mathematics.

The triplet, developed in 1882, of analysis, algebra, and arithmetic, is mirrored in the fact that the sine function is known by its zeros (at 0, $\pi$ and $2\pi$ and $3\pi$ and…); that sin(2x) may be expressed in terms of the sine function (=2 sin $x \times$ sin($\pi/2-x$)), what is called automorphy (having the same form); and, combinatorially by the fact that the sine function packages the odd factorials (namely, sin $x = \Sigma x^n/n!$ for odd n) and arithmetically sin (x+y) may be seen as a matter of adding numbers, actually the angles *x* and *y*, on a circle (rather than on a straight line). Similarly, there are curves described by a cubic equation (elliptic curves), which have nice pictures of their surfaces ala Riemann, they are functions defined algebraically, and one may do arithmetic on those curves.[7]

Another such example is found in how we understand the diffusion of a gas, such as the fragrance of a perfume, as someone enters the room and that fragrance is spread throughout that room: analytically, it can be represented by a differential equation; functionally and algebraically, diffusion spreads as √time, so that it takes four times as long to go twice the distance; and arithmetically, the diffusion of the fragrance molecules is a sum of random movements (each a molecular collision that leads to that diffusion).[8]

Ising Matter As An Identity In A Manifold Presentation Of Profiles

Ising matter would seem to allow for various ways of constituting it, and for computing its properties. From these ways, we might discern the following features:

Ising matter is composed of *particles*, not the molecules themselves, but coherently organized groups of molecules, those particles apparently acting much like electrons (fermions[9], spin ½ particles), each particle paired up with its sibling, what is called a Cooper pair (as in the theory of superconductivity). Those particles couple directly to the temperature field (rather than the molecules coupling to other magnetic molecules or external magnetic fields).

There are other "particles" that remind one of billiard balls scattering off each other, conserving energy and momentum, and when multiple particles are involved in a collision one might understand that collision as a set of two-body collisions. (This is known as the Yang-Baxter relation.)

Second, Ising matter is actually composed of *polymers*, and so we might study their paths through the lattice. We might make series approximations taking into account longer and longer such polymers or polygons.

Those polymers (as in *a* above) might well be interfaces (*d*) between clusters of molecules that are like-pointing, and one studies those interfaces rather than the conventional averaged bulk properties of Ising matter. This has proved fruitful close to or at the critical temperature.

Third, Ising matter is actually *blocks of blocks* of molecules from 2×2, to 2-(blocks of 2×2) ×2-(blocks of 2×2), and on the way up.

Fourth, Ising matter would seem to be much like a *random walk*, or a random field, one that looks the same as you coarse-grain it (combining groups of steps, looking with less resolution), with the only change indicated by a proportionality constant. But the random walks here are to some extent not so independent, and so that the Gaussian distribution for independent random variables is replaced by other distributions for strongly- and systematically-dependent random variables (which turn out to be much like the distribution of the eigenvalues of matrices with random entries).

Fifth, the Ising lattice, like crystal lattices, has *symmetries*—some of them are spatial: translational invariance, "duality" (a lattice and its dual, where points become lines, lines become points; some of them are dynamic: represented by $k$ and $u$; and some dynamic, mirrored by those spatial symmetries: duality, star-triangle. At the critical point, the lattice is the same as its dual, it is self-dual.

Sixth, the Ising lattice, like many a lattice or apparently-orderly system, goes from *order to disorder* as the temperature rises, with the feature that what happens at very high and very low temperatures mirror each other.

Seventh, the Ising lattice is a model of a discrete *quantum field theory*, of fermions, where the usual issues of analyticity, dispersion relations, etc., may play a role.

Various mathematical objects play a role in the methods of solution for the partition function, for the permanent magnetization, and for the relationship or correlation of spins far from each other.[10] What is also remarkable is how some of the earliest ideas in solving the Ising model become of great mathematical interest and complexity, especially in recent mathematically rigorous work on understanding the conformal[11] invariance of the Ising lattice at its critical point.[12]

Again, the main point here is that the Ising lattice, itself, may be seen and understood in manifold presentations or profiles or perspectives. Yet, they all are about that same lattice, that identity, and they are complementary even though they might well be very different in style and substance. Moreover, the mathematicians' work proves to be laden with physical meaning, and the physicists' work suggests and parallels interesting and important mathematics.

NOTES

[1]. Here I am trying to focus on the various qualities or characteristics (or "profiles") of Ising matter, each a consequence of one or another modes of modeling and calculation. The triplet I refer to has its most poignant description in a letter by André Weil, 1940, translated in the Appendix to Krieger, *Doing Mathematics*.

[2]. M. H. Krieger, *Doing Mathematics* (Singapore: World Scientific, 2015), chapters 3 and 5, provides a review of Ising model solutions and properties. His *Constitutions of Matter* (Chicago: University of Chicago Press, 1996) reviews some of the various models of matter. Various papers referred to here are in their bibliographies.

[3]. Onsager provided the first exact derivation in 1944, and there have been very many subsequent derivations. See note 2.

[4]. Analogously, the primes rather than the weighted number of states as in the partition function, are "packaged" by the zeta function, $\varsigma(s) = \Pi_{\text{over the primes}} (1-p^{-s})^{-1}$, called Euler products, which by unique factorization of the integers into primes is $\varsigma(s) = \sum_{\text{over the integers}} n^{-s}$ (its original definition), whose zeros would seem to be on a line in the complex plane and which exhibit other such features. But zeta does not reveal symmetries to be discovered when that zeta function is shown to be related to a theta function, an elliptic function, useful for describing heat flow in a metal slab (as Fourier showed, using what we now call Fourier or harmonic analysis), which exhibits lovely scaling properties and other symmetries but would not appear to have anything to do with the prime numbers. In this sense, the primes may be understood arithmetically and "automorphically" (having the same shape, as in scaling). Riemann connected zeta to the theta function ($\varsigma$ to $\theta$). The analytic (heat flow is smooth, for example) and algebraic features of theta allow one to better understand zeta (an apparently arithmetic or combinatorial function) and so the primes.

Similarly, the zeta function for a elliptic curve (a curve described by a cubic equation), called its L-function, packaging the number of solutions to that equation mod *m*, is the same as the L-function whose coefficients are taken from the fourier transform of an automorphic function (a modular form, that is, when you transform the independent variable, the function is changed by a multiplicative factor or modulus), the terms of art being that the "motivic" and the "automorphic" are in the end the same. But what you can learn from each L-function is different and complementary.

Another such averaging operator are the Hecke operators, $T_p$, averaging (over a group) the value of a modular function *f*. Those numbers are also available from the Galois representation, $M_p$, of an equation, as trace-$M_p$, for each prime *p*—and this brings to my mind that the partition function is the trace of the transfer matrix.

[5]. See S. Smirnov and D. Cheldak, "Discrete complex analysis on isoradial graphs," *Advances in Mathematics*, **228** (2011), 1590–1630.

[6]. R. Dedekind and H. Weber, "Theorie der algebraische Functionen einer Veränderlichen," Journal für reine und angewandte Mathematik **92** (1882):181-290. Translated in *Theory of Algebraic Functions in One Variable*, tr. J. Stillwell (Providence: American Mathematical Society, 2012).

[7]. See H. McKean and V. Moll, *Elliptic Curves: Function Theory, Geometry, Arithmetic* (Cambridge: Cambridge University Press, 1997).

[8]. More recently, Robert Langlands' program of connecting number theory to Fourier or harmonic analysis, eventually to be employed in Wiles' proof of the Fermat theorem, is echoed in the Ising model as well. (See for an introduction, R. Langlands, "Representation Theory: Its Rise and Role in Number Theory," in D.G. Caldi and G.D. Mostow, eds., *Proceedings of The Gibbs Symposium* (Providence: American Mathematical Society, 1990), pp. 181–210.)

The motivation is seen in the zeta function: again, what you can know about zeta depends on theta, that is, a connection of an arithmetic object with a function theoretic and analytic object, $f$ or $\theta$, an object that has lovely automorphic properties, again, namely $f(Rx)$ is nicely connected to $f(x)$, where $R$ is a rational linear transformation. We also note that the coefficients of the Fourier transform of $\theta^r$ are combinatorial numbers—the number of ways of forming a number from a sum of $r$ squares. Langlands connects an arithmetic object, the number of solutions to a polynomial Diophantine equation in a particular number system, to that polynomial's Galois group representation (a matrix equivalent), the Galois group reflecting some symmetries of that equation, and the traces or characters of that representation--*to* the Fourier coefficients of an automorphic object (a modular form and its "automorphic representation"). It is quite nontrivial to figure out just what should be the automorphic object for a particular equation. We connect the Galois group of the arithmetic problem *with* automorphic objects and their group representations.

The correspondence is expressed in terms of that multiplicative zeta-like L-functions, that multiplicative form called an Euler product. The L-function packaging the number of rational solutions to the elliptic curve (mod $p$) corresponds to an L-function packaging the Fourier coefficients of a modular form—or elevated on both sides to group representations and their traces. One can form an Euler product using the traces of the Galois group representation, $\pi$, to get $L_\pi$, or the Fourier coefficients of the modular form, $f$, to get $L_f$ and they should be the same. $L_\pi$ packages the numbers of interest. $L_f$ allows us to prove good smoothness properties and a functional equation, connecting $L_f$ at $s$ and at 1-$s$, to show analyticity in the complex-$s$ plane. ("Why Galois representations should be **the** source of Euler products with good functional equations is a complete mystery." R. Taylor, R., "Galois Representations," ms., arXiv:math/0212403, 2002, p. 13.)

More generally, Langlands' work is an extension of earlier work on quadratic reciprocity (Gauss to at least Artin), a way of determining whether a number is prime, and more generally of factoring polynomials. Moreover, we might think of the prime numbers and the elementary particles as playing similar roles in their respective disciplines.

[9]. A fermion, like an electron, is characterized by the Pauli Exclusion Principle: there can be only one of each distinct set of properties. Put differently, in walking around you cannot backtrack, for then your path would have two passages over the same line.

[10]. As for the various mathematical objects: Painlevé transcendents (distant cousins of sines and cosines); matrices with random entries; various universal distributions beyond the Gaussian (Wigner, Tracy-Widom); the highly-symmetric Toeplitz matrices, and Wiener-Hopf methods used to solve them for the permanent magnetization (in effect accounting for the effects of the molecules upon each other, adding up their contribution to the magnetization); elliptic functions, elliptic integrals, elliptic curves; and functional equations. Moreover, there are remarkable features in the series expansions of the correlation functions, involving various differential operators, and features of elliptic curves.

[11]. Conformal invariance is the fact that one might stretch a two-dimensional rubber sheet in many ways yet its most fundamental characteristics remain the same.

[12]. Smirnov's work (see note 5 above) involves Kac-Ward determinants, Kramers-Wannier duality, fermionic observables (presumably related to the fermions I refer to above), isoradial graphs (that in effect have all the good properties of square graphs, that is, duality).